\documentclass[final]{cvpr}

\usepackage{times}
\usepackage{epsfig}
\usepackage{graphicx}
\usepackage{amsmath}
\usepackage{amssymb}
\usepackage{amsmath,amsfonts,xcolor,graphicx,bm, adjustbox, tabularx}
\usepackage{float}
\usepackage{dblfloatfix}
\usepackage{color-edits}
\newcommand{\R}{\mathbb{R}}
\usepackage[pagebackref=true,breaklinks=true,colorlinks,bookmarks=false]{hyperref}
\usepackage{enumitem}
\setlist{nolistsep}

\begin{document}

\title{Noise2Inpaint: Learning Referenceless Denoising by Inpainting Unrolling}
	\author{Burhaneddin Yaman\textsuperscript{\rm 1,2,*} \qquad Seyed Amir Hossein Hosseini\textsuperscript{\rm 1,2,}\thanks{First two authors contributed equally to this work.} \qquad Mehmet Ak\c{c}akaya\textsuperscript{\rm 1,2}\\
	
		{\textsuperscript{\rm 1} Department of Electrical and Computer Engineering, University of Minnesota, Minneapolis, MN, USA}  \\
		{\textsuperscript{\rm 2}Center for Magnetic Resonance Research, University of Minnesota, Minneapolis, MN, USA} \\
		\small{\texttt{\{yaman013, hosse049, akcakaya\}@umn.edu}}

	}
\maketitle

\begin{abstract}
Deep learning based image denoising methods have been recently popular due to their improved performance. Traditionally, these methods are trained in a supervised manner, requiring a set of noisy input and clean target image pairs. More recently, self-supervised approaches have been proposed to learn denoising from only noisy images. These methods assume that noise across pixels is statistically independent, and the underlying image pixels show spatial correlations across neighborhoods. These methods rely on a masking approach that divides the image pixels into two disjoint sets, where one is used as input to the network while the other is used to define the loss. However, these previous self-supervised approaches rely on a purely data-driven regularization neural network without explicitly taking the masking model into account. In this work, building on these self-supervised approaches, we introduce Noise2Inpaint (N2I), a training approach that recasts the denoising problem into a regularized image inpainting framework. This allows us to use an objective function, which can incorporate different statistical properties of the noise as needed. We use algorithm unrolling to unroll an iterative optimization for solving this objective function and train the unrolled network end-to-end. The training paradigm follows the masking approach from previous works, splitting the pixels into two disjoint sets. Importantly, one of these is now used to impose data fidelity in the unrolled network, while the other still defines the loss. We demonstrate that N2I performs successful denoising on real-world datasets, while better preserving details compared to its purely data-driven counterpart Noise2Self.
\end{abstract}

\section{Introduction}
 Image denoising aims to recover clean images from noisy measurements, since it is not feasible to avoid noise contamination in numerous scenarios due to instrumental imperfection or environmental conditions.  
The implicit assumption of many denoising approaches is that pixels of the underlying clean images are spatially correlated, while the contaminating noise instances are uncorrelated \cite{NonLocalMeans,BM3D_dabov2007,Denosing_elad2006image,Denoising_zoran2011learning}. In recent years, convolutional neural networks (CNNs) have gained immense attention for image denoising \cite{zhang2017learning,RED,Ulyanov_2018_CVPR,N2V,Soltanayev_SURE,N2N,milanfar_denoising}. In CNN-based denoising, parameters of the convolutional kernels are traditionally tuned to minimize the discrepancy between pairs of noisy and clean target images as measured by a pre-specified loss metric \cite{zhang2017learning,lefkimmiatis2018universal}. The network is consequently expected to generalize to denoise future images with similar statistical properties.

The classical supervised training of CNN-based denoiser requires availability of clean target images pertinent to noisy ones, which may not be readily available in some scenarios \cite{N2N,N2V,N2S}. A number of recent research studies have attempted to address this issue by training CNNs without requiring ground truth. Noise2Noise \cite{N2N} was the first method  that proposed to perform the training on pairs of noisy images rather than noisy and clean images. The main underlying assumption of Noise2Noise is the availability of two noisy instances of the same image with independent noise, which may be difficult to obtain in practice, such as medical imaging applications. 

To tackle this challenge, several self-supervised approaches have been proposed to learn denoising from only noisy images \cite{N2V,N2S,self2self,Noise2Same}. The main underlying assumptions in these approaches are the statistical independence of noise across pixels, and the existence of spatial correlations across the pixels of the true underlying image. These self-supervised approaches split image pixels into two disjoint sets following a masking operation, in which image pixels in one of these set is used as input to the network while the other is used to define the loss. Among such self-supervised approaches, Noise2Self (N2S) theoretically shows that under a certain masking choice, minimizing the self-supervised loss on only noisy images is equivalent to minimizing the supervised loss function up to a constant under the aforementioned assumptions. While self-supervised approaches learn denoising using only noisy images, all these approaches relies on a purely data-driven regularization neural network without explicitly incorporating the masking model in the network architecture.

In this work, we propose a novel self-supervised image denoising approach, called Noise2Inpaint. Noise2Inpaint utilizes the masking model of N2S, but recasts the denoising problem as an image inpainting inverse problem with a well-defined objective function. Subsequently, an iterative optimization procedure for solving this objective function is unrolled \cite{gregor2010learning}, incorporating a CNN-based regularizer and a linear data fidelity unit in each iteration \cite{pdCTbyOktem,hammernik2018learning,deepADMMnet}. The latter allows N2I to explicitly use the masking model in the network architecture. The masking model of N2I follows N2S, splitting the noisy pixels into two disjoint sets. Notably, one of these sets is now used in the data fidelity units in the unrolled network in contrast to previous purely data-driven works, while the other is similarly used to define the loss function.

Our  main contributions are summarized as follows: 
\begin{itemize}
\item Introducing a self-supervised learning approach for referenceless denoising by recasting the denoising problem as an inpainting task with an objective function to be minimized in an optimization framework.
\item Training an unrolled neural network with several data fidelity and CNN-based regularization units to solve the optimization problem pertinent to the inpainting challenge using self-supervision. 
\item Providing an objective function that can incorporate different noise statistics, which can be used to generalize the approach to different noise models either during training or testing.
\item{Applying the proposed Noise2Inpaint approach to real world datasets, and showing its superiority compared to its purely data-driven counterparts.}
\end{itemize}

\begin{figure*}[!t]
    \begin{center}
            \includegraphics[width=0.9\textwidth]{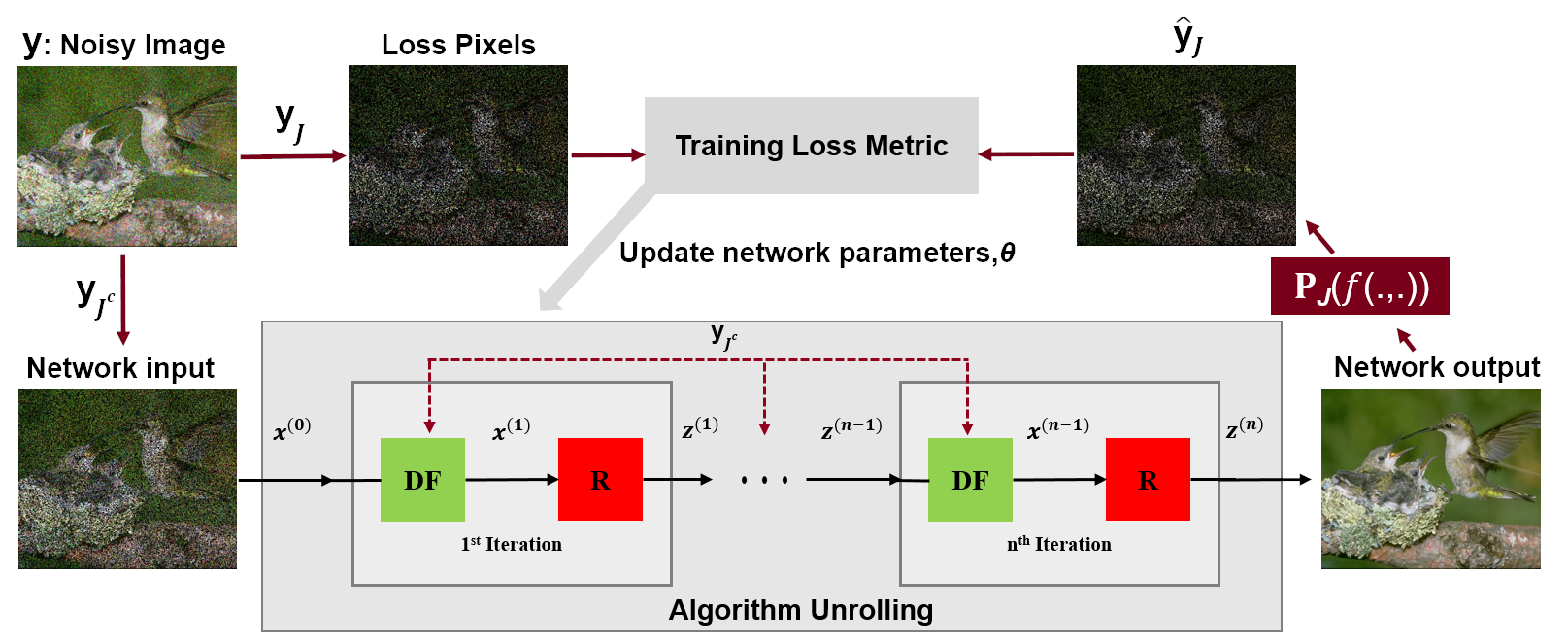}
    \end{center}
    \vspace{-.3cm}
    \caption 
    {\label{fig:algorithm unrolling} The self-supervised training in Noise2Inpaint for training the unrolled network to solve equations in \ref{Eq:recons3a} and \ref{Eq:recons3b}. Each step in the unrolled network consist of data fidelity (DF) and regularizer unit. } 
    \vspace{-.25cm}
\end{figure*} 

\section{Related Work} \label{sec:related works}
In this section, we discuss CNN-based denoising algorithms, including approaches that do not require ground truth clean data for training.

Image denoising focuses on recovering a clean target image, $\mathbf{x} \in \mathbb{R}^m$  from a noisy image, ${\bf y} \in \mathbb{R}^m$. Typically, an additive noise model is assumed \cite{BM3D_dabov2007,Denosing_elad2006image} with
\begin{equation}\label{Eq:Denoising}
   \bf  y = x + n,
\end{equation}
where $\bf n$ denotes the noise which is generally modeled as Gaussian and is independent from $\bf x$ \cite{NonLocalMeans,BM3D_dabov2007}.
More complicated statistical models may also be encountered in practical applications \cite{Rician_liu2010denoising,Rician_manjon2012new,DenoisingOverview_Siam,DenoisingOverview}. 
\subsection{Noise2True Training}
A common setup for deep learning methods in image denoising is the supervised setting, which is also referred to as Noise2True (N2T), and requires clean ground-truth images for training. Specifically, N2T training learns a  denoising CNN $f: \R^m \to \R^m$ that is parameterized by $\bm \theta$ and which minimizes the supervised loss
\begin{equation}\label{eqn:supervised_loss}
    \mathcal{L}({\bm \theta}) \triangleq \mathbb{E} \|f({\bf y};{\bm \theta}) - {\bf x}\|^2.
\end{equation}
Denoising function $f$ in Eq. \ref{eqn:supervised_loss} is approximated by minimizing the empirical loss on  a database of pairs of noisy input and ground truth clean images $\{{\bf y}^i, {\bf x}^i\}_{i=1}^N$ as  
\begin{equation}
   \min_{\bm \theta} \mathcal{L}_N ({\bm \theta})\triangleq \sum_{i=1}^N  \|f({\bf y}^i; {\bm \theta}) - {\bf x}^i \|^2,
   \label{Eq:Sup}
\end{equation}
where $f(.;{\bm \theta})$ is the output of the CNN with parameters ${\bm \theta}$, and $\mathcal{L}_N({\bm \theta})$ is the empirical loss function \cite{zhang2017learning,Denoising_CNN_zhang2018ffdnet,N2N,N2S}. 

\subsection{Noise2Noise Training}
Noise2Noise (N2N) training does not require any clean images \cite{N2N} as opposed to N2T. N2N performs training by learning a mapping function between pairs of noisy images that have the same underlying clean images but independently drawn noises from the same distribution. 
The key concept of N2N is that given a set of two such degraded images,
$\{ {\bf y}^i =  {\bf x}^i+n^i, {\bf \hat{y}}^i =  {\bf x}^i+\hat{n}^i\}$, the expected value of both of these noisy images are equivalent to the clean signal. Hence, N2N modifies loss function in  Eq. \ref{Eq:Sup} into 

\begin{equation}
   \min_{\bm \theta} \sum_{i=1}^N  \|f({\bf y}^i; {\bm \theta}) - {\bf \hat{y}}^i \|^2.
   \label{Eq:N2N}
\end{equation}

\subsection{Noise2Self Training} \label{sec:N2V-N2S}
Self-supervised training approaches enable training of CNNs without requiring a ground-truth image or pairs of noisy images. Under the assumption of independent zero-mean noise across pixels, these methods minimize a self-supervised loss on noisy images as
\begin{equation}\label{eqn:self_loss}
    \mathcal{L}({\bm \theta}) \triangleq \mathbb{E} \|f({\bf y};{\bm \theta}) - {\bf y}\|^2.
\end{equation}

The early pioneering work in this field is Noise2Void (N2V), which takes random patches from images and replaces the central pixel with a random pixel in the patch. Training is performed by minimizing a loss function between only the true central pixel value and estimated central pixel value of the network. While N2V empirically works well, it lacks theoretical guarantees as Eq. \ref{eqn:self_loss}  simplifies to learning an identity mapping.  Noise2Self (N2S) provides strong theoretical guarantees by proposing a $\mathcal{J}$-invariant $f$ that avoids learning the identity function \cite{N2S}. 

\noindent\textbf{Definition.} Let $\mathcal{J}=\{J_1,\dots, J_N\}$ be a set of partitions of the pixels set of an image, where $\sum_{i=1}^N |J_i|=m$. A function  $f: \R^m \to \R^m$ is $\mathcal{J}$-invariant if the value of $f(\mathbf{y})_J$ does not depend on the value of $ \mathbf{y}_J$ for all $J\in\mathcal{J}$ \cite{N2S}.

\noindent In other words, the pixel values of an image is split into two disjoint sets $J$ and $J^c$ with $|J|+|J|^c=m$, and denoising function $f({\bf y})_J$ uses pixels in ${\bf y}_{J^c}$ to denoise ${\bf y}_J$.  Hence, rewriting the self-supevised loss function in Eq. \ref{eqn:self_loss} over $\mathcal{J}$-invariant functions leads to \cite{N2S}
\begin{equation}\label{eqn:n2self_loss}
      \mathbb{E} \|f({\bf y};{\bm \theta}) - {\bf y}\|^2 =  \mathbb{E} \|f({\bf y};{\bm \theta}) - {\bf x}\|^2 + \|{\bf y} - {\bf x}\|^2 .  
\end{equation}
Thus, minimizing the self-supervised loss over $\mathcal{J}$-invariant function is equivalent to minimizing supervised loss up to a constant defined by the variance of noise (last term). Hence, \noindent $\mathcal{J}$-invariant denoising function $f$ can be empirically approximated over a database of noisy images as
\begin{equation}
     \arg \min_{\bm \theta}  \sum_{i} \sum_{J \in \mathcal{J}} \|{\bf P}_{J}f({\bf y}_{J^c}^i; {\bm \theta}) - \:{\bf y}_{J}^i \|^2,
\end{equation}
where ${\bf P}_{J}$ is defined as the masking operator specified by the index set $J$ in order to perform the loss. 

\section{Methods}  \label{sec:methods}
Our proposed Noise2Inpaint (N2I) method builds on the $\mathcal{J}$-invariant masking idea from N2S \cite{N2S}, as well as a recent self-supervision method from image reconstruction that uses an optimization-focused algorithm unrolling for neural networks \cite{yaman_SSDU_MRM}. In conventional image denoising, a regularized objective function is typically used \cite{RED}: 
\begin{equation}
\label{Eq:recons1}
\arg \min_{\bf x} \|\mathbf{y}-\mathbf{x}\|^2_2 + \cal{R}(\mathbf{x}),
\end{equation}
where the first term denotes a data fidelity term between the desired output and the noisy input, while the second term $\cal{R}(\mathbf{\cdot})$ is a regularizer. These regularizers can have explicit closed forms \cite{Pock_TGV,donoho1995noising}, or the whole objective function can be solved implicitly either with traditional methods \cite{NonLocalMeans,BM3D_dabov2007} or using CNNs \cite{Denoising_CNN_gharbi2016deep,Ulyanov_2018_CVPR,Denoising_CNN_Xu_2015_ICCV,zhang2017learning,Denoising_CNN_zhang2018ffdnet}.

Here, instead of viewing the masking approach in N2S as estimating a pixel in $J$ from its complement $J^c$, we recast it as an image inpainting problem \cite{Inpainting_Overview}. 
In image inpainting, missing pixels are estimated from available pixels using a regularized objective function. Similar to N2S, let $J$ be the masked pixels, and $J^c$ be the complement pixels that are available at the input of the neural network. Then, the available non-masked data in Eq. (\ref{Eq:Denoising}) is given as 
\begin{equation}\label{Eq:Inpainting}
   {\bf  y}_{J^C} = {\bf P}_{J^C}{\bf x} + {\bf n}_{J^C}
\end{equation}
where ${\bf P}_{J^C}$ is the masking operator as defined in Section \ref{sec:N2V-N2S}. While inpainting seems like a more difficult problem than denoising, this recasting allows us to write an objective function that can be solved using algorithms that enforce data fidelity and regularization.

\subsection{Algorithm Unrolling for Inpainting}
The objective function corresponding to the measurement model in Equation (\ref{Eq:Inpainting})
for the inpainting problem is given as
\begin{equation}
\label{Eq:recons2}
\arg \min_{\bf x} \|{\bf  y}_{J^C} - {\bf P}_{J^C}{\bf x}  \|^2_2 + \cal{R}(\mathbf{x}),
\end{equation}

The regularized inpainting problem has been extensively studied using only CNNs \cite{Inpainting_Pathak2016,Inpainting_ShiftNet,Inpainting_Yang_HighRes2017}. An alternative approach to solve the regularized inpainting problem is to use algorithm unrolling \cite{gregor2010learning}. In these methods, an iterative optimization algorithm, such as proximal gradient descent or variable splitting \cite{Proximal_combettes2011proximal,VS_wang2008new} for solving the objective function in Equation (\ref{Eq:recons2}) is unrolled for a fixed number of iterations. Each iteration consists of a data fidelity and regularizer term, as shown in  Figure \ref{fig:algorithm unrolling}. The unrolled network is trained end-to-end by minimizing a loss function that characterizes the discrepancy between a reference and network output. Algorithm unrolling has gained significant popularity in many fields such as image reconstruction tasks in MRI or computational tomography due to its improved precision and accuracy \cite{hammernik2018learning,Zhussip_2019_CVPR,kellman2019physics,sreehari2016plug,yang2018low,yaman_SSDU_MRM}.

One approach to solve the objective function in Equation (\ref{Eq:recons2}) is to use variable splitting (VS) \cite{VS_wang2008new}. 
VS decouples the data fidelity and regularizer term by introducing an auxilliary variable $\mathbf{z}$ that is constrained to be equal to $\mathbf{x}$.  Following variable spliting approach, the objective function in Equation (\ref{Eq:recons2}) can be rewritten using quadratic relaxation:
\begin{equation}
\label{Eq:recons3}
\arg \min_{\bf {x, z}}\|{\bf  y}_{J^C} - {\bf P}_{J^C}{\bf x}\|^2_2 + \mu \lVert\mathbf{x}-\mathbf{z}\rVert_{2}^2 +\cal{R}(\mathbf{z}),
\end{equation}
where $\mu$ denotes the penalty parameter. This is solved by alternating minimization over $\mathbf{x}$ and $\mathbf{z}$ as
\begin{subequations}
\begin{align}
& {\bf x}^{(k)} =\arg \min_{\bf x} \|{\bf  y}_{J^C} - {\bf P}_{J^C}{\bf x}\|^2_2 + \mu \lVert{\bf x}-{\bf z}^{(k-1)}\rVert_{2}^2 \label{Eq:recons3a}
\\
& {\bf z}^{(k)} = \arg \min_{\bf z}\mu \lVert {\bf x}^{(k)}-{\bf z}\rVert_{2}^2 +\cal{R}({\bf z})\label{Eq:recons3b}
\end{align}
\end{subequations}

In algorithm unrolling, this problem is unrolled for a fixed number of iterations, with each iteration including a data fidelity and a regularization block. The regularization subproblem in Equation (\ref{Eq:recons3b}) does not have a closed form solution and is solved implicitly by CNNs. Equation (\ref{Eq:recons3a}) corresponds to the data fidelity term, with a closed form solution 
\begin{equation}
\mathbf{x}_j^{(k)} = 
\begin{cases}
  \mathbf{z}_j^{(k-1)} & \text{if } j \in J \\
  \frac{1}{1+\mu}{\bf y}_j + \frac{\mu}{1+\mu}{\bf z}_j^{(k-1)} & \text{if }  j \in J^c
\end{cases}
\end{equation}
where $j$ indicates the pixel location in the image. In other words, at iteration $k$ in the unrolled network, the denoised image is comprised of the CNN output at the masked locations and a weighted average for the non-masked locations.

\subsection{Noise2Inpaint Self-Supervised Training}
The proposed Noise2Inpaint method performs end-to-end training by minimizing 

\begin{equation} \label{ssdu:train}
    \arg \min_{\bm \theta}  \sum_{i} \sum_{J \in \mathcal{J}} \|{\bf P}_{J}\big(f_\textrm{unroll}({\bf y}_{J^c}^i,{\bf P}_{J^c}; {\bm \theta})\big) - {\bf y}_{J}^i  \|^2,
 \end{equation}
where $f_\textrm{unroll}({\bf y}_{J^c}^{i}, {\bf P}_{J^c}^{i}; {\bm \theta})$ denotes the output of the unrolled network for inpainting described by Equations (\ref{Eq:recons3a})-(\ref{Eq:recons3b}), with ${\bf y}_{J^c}^{i}$ and ${\bf P}_{J^c}^{i}$ denoting the inputs used at the data fidelity units of the unrolled network. $\bm \theta$ includes the parameters of the CNN that implements the regularization unit of Equation (\ref{Eq:recons3b}), as well as the learnable quadratic penalty parameter $\mu$ used in Equation (\ref{Eq:recons3b}).

We note a few important points about the objective function in Equation (\ref{ssdu:train}): First, the loss is defined between noisy pixels excluded in the training and the network output at corresponding unseen locations, which is reminiscent of N2S. Second, in contrast to  N2S, the N2I network has a well-defined separation between linear data consistency and the CNN-based regularization units. Third, as the weights of the regularization CNNs are shared across the iterations of the unrolled network, N2I has only one additional parameter (the penalty term $\mu$) compared to N2S when using the same CNN architecture. Finally, when using the masking scheme described in \cite{N2S} for selecting $J$, the N2I enjoys the same theoretical guarantees as N2S.

\section{Experiments}
\label{sec:experiments}

The proposed Noise2Inpaint method is evaluated on various denoising tasks. We compare our results with Noise2True, Noise2Noise, Noise2Self and a conventional denoising algorithm BM3D \cite{dabov2007image}. 
\begin{figure*}[t]
    \begin{center}
            \includegraphics[width=1\textwidth]{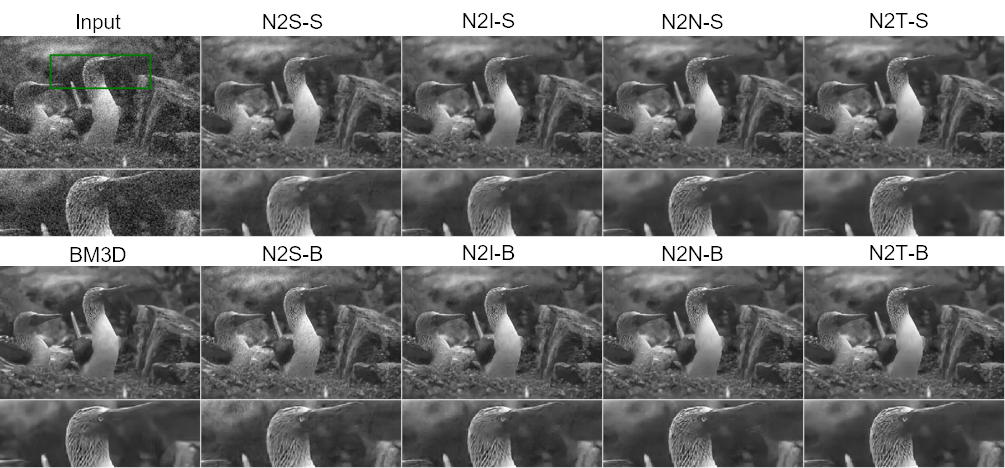}
    \end{center}
    \vspace{-.3cm}
    \caption 
    {\label{fig:BSD_fig} Denoising results of one representative image from BSD68 with noise level 25 for training with known specific (-S) and blind (-B) Gaussian noise. (N2S: Noise2Self, N2I: Noise2Inpaint (Ours), N2N: Noise2Noise, N2T: Noise2Truth)}
    \vspace{-.25cm}
\end{figure*}

\begin{table*}[b]
\begin{adjustbox}{width=0.9\textwidth,center}
\begin{tabular}{|l|l|l|l|l|l|l|l|l|l|}
\hline
Methods     & BM3D  & N2S-S & N2S-B & N2I-S & N2I-B & N2N-S & N2N-B & N2T-S & N2T-B \\ \hline
$\sigma$=15 & 31.09 & 25.54 & 27.31 & 30.14 & 30.1  & 31.21 & 30.74 & 31.23 & 30.85 \\ \hline
$\sigma$=25 & 28.59 & 26.51 & 25.64 & 28.02 & 28.03 & 28.72 & 28.35 & 28.75 & 28.49 \\ \hline
$\sigma$=50 & 25.68 & 23.55 & 23.58 & 25.17 & 24.92 & 25.75 & 25.31 & 25.75 & 25.56 \\ \hline
\end{tabular}
\end{adjustbox}
\vspace{.1cm}

  \caption{Average PSNR results on the BSD68 dataset for known specific and blind Gaussian denoising using BM3D, Noise2Self-Specific/Blind(N2S-S,N2S-B), Noise2Inpaint-Specific/Blind (N2I-S,N2I-B), Noise2Noise-Specific/Blind (N2N-S,N2N-B) and Noise2True-Specific/Blind (N2T-S,N2T-B). 
  }
  \label{tbl:BSD_table}
\end{table*}

\subsection{Datasets}

\noindent\textbf{BSD.} A grey-scale natural images dataset is generated from the Berkeley Segmentation Dataset (BSD) \cite{BSDDatabase}, following \cite{DnCNN}. This dataset contains 400 cropped images of size 180$\times$180 with a pixel intensity range of [0-255]. BSD68 dataset (68 grey-scale images) is used for testing.

\noindent\textbf{H\`{a}nz\`{i}.}  We construct a dataset of 13029 Chinese characters (H\`{a}nz\`{i}). As in \cite{N2S}, the whole dataset comprises of 78174 images (each character is repeated 6 times) of size 64 $\times$ 64 and a pixel intensity range of [0-1]. The dataset is split into two as 90$\%$ and 10 $\%$ for training and testing, respectively. 

\noindent\textbf{ImageNet.} To generate a dataset of RGB natural images, ImageNet LSVRC 2012 Validation dataset consisting of 50,000 images is used \cite{N2S,Noise2Same}. The training dataset of 60,000 cropped images of size 128 $\times$ 128 wih a pixel intensity range of [0-255] is constructed from the first 20,000 images. Another 1,000 different images are used for testing.

\noindent\textbf{Fluorescence Microscopy.} To show the utility of self-supervised denoising methods in real world applications, two fluorescence microscopy datasets from the Cell Tracking challenge \cite{CellTrack_ulman2017objective} (Fluo-C2DL-MSC and Fluo-N2DH-GOWT1) are used. These datasets only contain single noisy images \cite{N2V}. Training dataset for each of these microscopy datasets contain 100 cropped images of size 512 $\times$ 512 with a pixel intensity range of [0-255]. 
  
\subsection{Implementation Details}
Experiments are performed on two U-Net architectures. We use a shallow U-Net architecture of depth 2, with a linear function in the last layer \cite{N2V} for all experiments on BSD and fluorescence microscopy datasets. For the experiments on Hanzi and ImageNet datasets, we use a deeper U-Net (depth 4) architecture with batch normalization and a batch size of 64 as in \cite{N2S}. For both networks: The number of channel in initial level is set to 32 channels and it doubles as it goes deeper; kernel size 3; Adam optimizer with learning rate of $10^{-5}$. Since our study focuses on enabling self-supervised learning from an algorithm unrolling perspective, we use CNNs that have been previously utilized in self-supervised denoising literature. However, further improvements may be possible with other well-designed neural networks. PSNR is used as evaluation criterion, when a reference image is available. Training datasets are augmented by rotating each image 90$^{\circ}$ three times and and mirroring them. 

Noise2True, Noise2Noise and Noise2Self are trained as described in Section \ref{sec:related works}. These methods use a purely data-driven CNN as previously described. Our Noise2Inpaint method is trained end-to-end by unrolling the iterative algorithm in Eq. \ref{Eq:recons3a}, and \ref{Eq:recons3b} for 10 iterations. Each iteration contains a data fidelity and regularization term, where the trainable parameters are shared across iterations. Hence, Noise2Inpaint has only 1 more trainable parameter, which is the $\mu$ penalty parameter for quadratic relaxation, compared to N2T, N2N and N2S. During training, we follow \cite{N2S} and randomly choose one single mask $J$ for each image with density $1/25$ to speed up the training process. At inference time, we input the full noisy image on the trained network as this has been reported \cite{N2S} to outperform the strategy of applying a partition $\cal{J}$ containing $|\cal{J}|$ sets and averaging them.

\begin{figure*}[t]
    \begin{center}
            \includegraphics[width=1\textwidth]{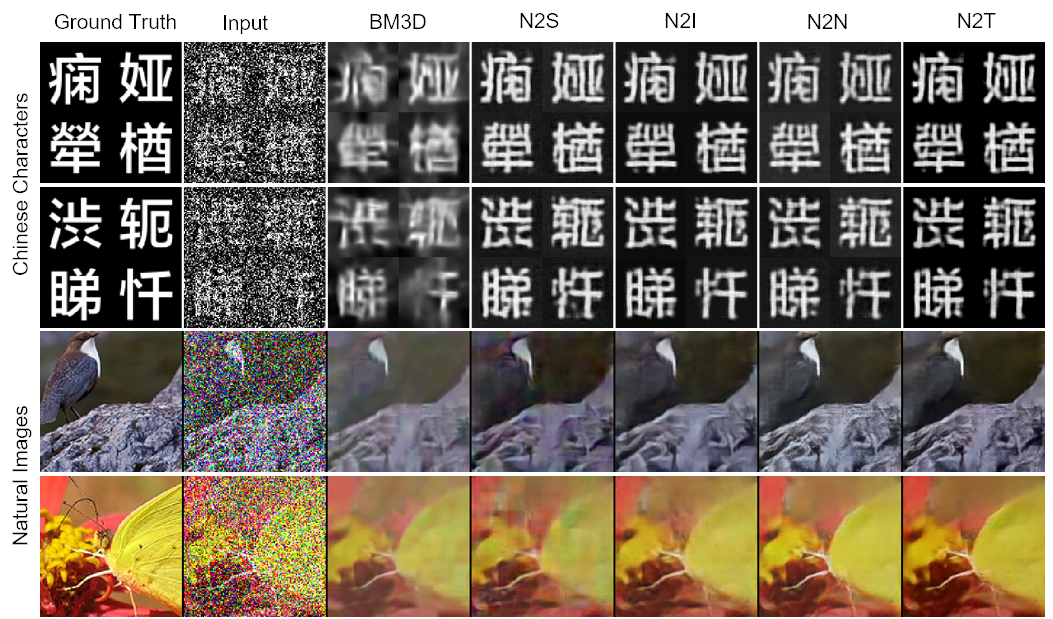}
    \end{center}
    \vspace{-.3cm}
    \caption 
    {\label{fig:Hanzi_ImageNet} Denoising performance on Chinese characters (H\`{a}nz\`{i}) and RGB natural images (ImageNet) test datasets using traditional denoising method BM3D, supervised method Noise2True, supervised with second noisy image Noise2Noise, self-supervised approaches Noise2Self and Noise2Inpaint.}
    \vspace{-.25cm}
\end{figure*} 

\subsection{Known and Blind Gaussian Noise Removal}
\label{sec:bsd_experiments}
We perform two set of experiments using BSD400 dataset to analyze denoising performance in the presence of known and blind Gaussian noise. For known noise level, we use three noise levels $\sigma = $15, 25 and 50, and train a network for each noise level \cite{DnCNN}. For blind Gaussian denoising, we train a single model using noisy images selected from noise levels $\sigma \in [0,50]$.

Average PSNR results on test BSD68 dataset for known and blind Gaussian denoising are shown in Table \ref{tbl:BSD_table}. We refer to each method trained with known specific and blind noise level as Method-S and Method-B, respectively following the convention of \cite{DnCNN}. As anticipated, Noise2True and Noise2Noise with known noise levels achieves the best PSNR results as they use extra information. Among the self-supervised approaches, our method Noise2Inpaint outperforms Noise2Self for both known and blind noise cases. Figure \ref{fig:BSD_fig} displays denoising results at noise level $\sigma$ = 25. The proposed Noise2Inpaint approach shows superior reconstruction quality compared to Noise2Self by preserving more details for known noise removal, and reducing non-uniform background artifacts with subjectively appealing reconstruction in blind denoising.

\begin{table}[b]
\begin{tabular}{|l|l|l|l|l|l|}
\hline
Methods  & BM3D  & N2S   & N2I   & N2N   & N2T   \\ \hline
Hanzi    & 10.69 & 12.70 & 13.55 & 12.79 & 13.99 \\ \hline
ImageNet & 18.18 & 19.12 & 20.26 & 20.72 & 20.97 \\ \hline
\end{tabular}
\vspace{.1cm}

  \caption{Average PSNR results on the H\`{a}nz\`{i} and ImageNet dataset for mixture noise levels. 
  } 
  \label{tbl:Hanzi_table}
\end{table}
\subsection{Mixture Noise Removal}
H\`{a}nz\`{i} and ImageNet datasets are evaluated with a mixture of different noise models. For the H\`{a}nz\`{i} dataset, a mixture of Gaussian ($\sigma$ = 0.7) and Bernoulli noise (half the pixels blacked out) are applied to each clean image, as in \cite{N2S}. For the ImageNet dataset, multiplicative Poisson noise ($\lambda$ = 30), additive Gaussian noise ($\sigma$ = 80) and Bernoulli noise ($p$ = 0.2) is applied to each clean image, following \cite{N2S}. 

\begin{figure}[!t]
\begin{center}
   \includegraphics[width=1\linewidth]{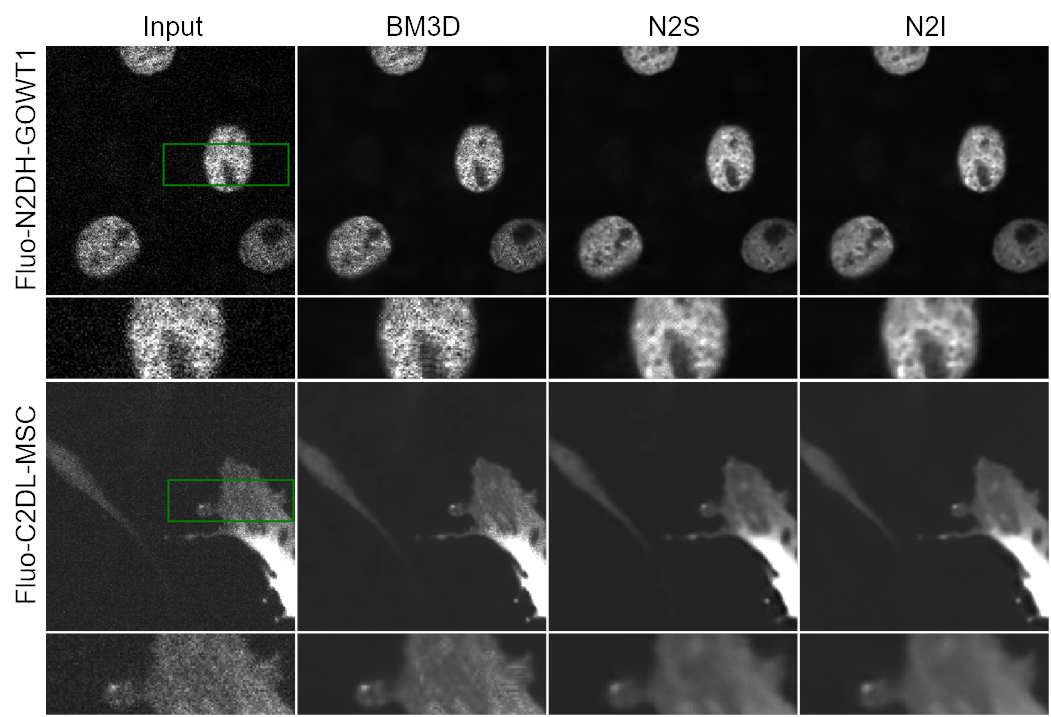}
\end{center}
  \caption 
    {\label{fig:Fluo} Representative results from fluorescence microscopy datasets Fluo-N2DH-GOWT1 and Fluo-C2DL-MSC for traditional denoising method BM3D and self-supervision methods Noise2Self and Noise2Inpaint. Note that Noise2True and Noise2Noise are not applicable as microscopy datasets contain only single noisy images.}
\end{figure}

\begin{figure*}[t]
    \begin{center}
      \includegraphics[width=1\textwidth]{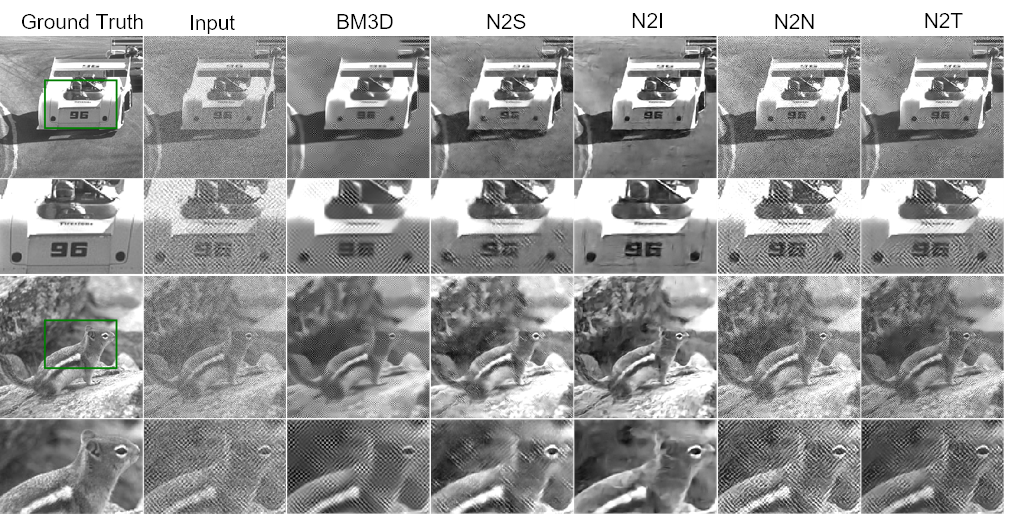}
    \end{center}
    \vspace{-.3cm}
    \caption 
    {\label{fig:colored_noise} Denoising results on BSD68 dataset in presence of spatially correlated noise. Proposed Noise2Inpaint approach removes the structured noise. BM3D and purely data-driven denoising methods(N2S/N2N/N2T) fails to remove structured noise.}
    \vspace{-.25cm}
\end{figure*}

Average PSNR values over the test datasets for both H\`{a}nz\`{i} and ImageNet are listed in Table \ref{tbl:Hanzi_table}. While N2T achieves the highest metrics among all methods, Noise2Inpaint achieves higher PSNR compared to the other ground-truth free approaches, N2S and BM3D. Figure \ref{fig:Hanzi_ImageNet} illustrates the visual results on representative test images. N2I achieves a better denoising quality compared to N2S and BM3D, which aligns with the quantitative metrics. 

\subsection{Denoising of Fluorescence Microscopy Data}
We evaluate performance of blind denoising methods on Fluorence microscopy datasets, which contain only single sets of noisy images. Hence, Noise2True and Noise2Noise are not applicable. Figure \ref{fig:Fluo} shows the performance of BM3D, and self-supervised N2S and N2I approaches. We assess the results qualitatively with visual inspection as quantitative metrics such as PSNR cannot be reported due to lack of ground-truth reference data. The zoomed-in regions show that Noise2Inpaint achieves a superior denoising quality compared to BM3D and N2S by suppressing the noise further and achieving a more spatially uniform and visually appealing result. 

\subsection{Structured Noise Removal} \label{sec:46}
Structured noise is a commonly encountered noise type especially in biomedical imaging applications \cite{goyal2018noise, Pruessmann1999}. However, Noise2Self and other referenceless denoising learning approaches have an independence assumption, which is violated with structured noise such as correlated Gaussian noise. On the contrary, the proposed Noise2Inpaint approach can effectively deal with structured noise, since the data fidelity term in the objective function in Eq. \ref{Eq:recons1}
can be recast as the log-likelihood term for different noise statistics as:
\begin{equation}
\label{Eq:log-liklihood1}
\arg \min_\mathbf{x} \: \textrm{-}\log(p(\mathbf{y}|\mathbf{x})) + \cal{R}(\mathbf{x}),
\end{equation}
where $p(\mathbf{y}|\mathbf{x})$ defines the likelihood of $\mathbf{y}$ given $\mathbf{x}$. Similarly, Eq. \ref{Eq:recons2} can be recast as
\begin{equation}
\label{Eq:log-liklihood}
\arg \min_\mathbf{x} \: \textrm{-}\log(p(\mathbf{y}_{J^C}|\mathbf{x})) + \cal{R}(\mathbf{x}),
\end{equation}
where $p(\mathbf{y}_{J^C}|\mathbf{x})$ defines the likelihood of $\mathbf{y}_{J^C}$ given $\mathbf{x}$. For instance, for colored Gaussian noise, Eq. \ref{Eq:log-liklihood} leads to 
\begin{equation}
\label{Eq:recons2-colored}
\arg \min_\mathbf{x} \|\textbf{K}_{J^C}^{-1/2}(\mathbf{y}_{J^C} - \mathbf{P}_{J^C}\mathbf{x})\|^2_2 + \cal{R}(\mathbf{x}),
\end{equation}
where $\textbf{K}_{J^C}$ is the covariance matrix of the noise vector defined over locations $J^C$. 

\begin{table}[b]
\begin{tabular}{|l|l|l|l|l|l|}
\hline
Methods  & BM3D  & N2S   & N2I   & N2N   & N2T   \\ \hline
BSD68    & 17.15 & 20.91 & 22.50 & 17.96 & 17.80 \\ \hline

\end{tabular}
\vspace{.1cm}

  \caption{Average PSNR results on the BSD68 dataset for structured noise. 
  } 
  \label{tbl:structured_noise}
\end{table}
By applying the variable splitting approach, the following sub-problems are alternatively solved over $\bf x$ and $\bf z$:
\begin{subequations}
\begin{align}
    & \mathbf{x}^{(k)} =\arg \min_\mathbf{x} \|\mathbf{K}_{J^C}^{-1/2}(\mathbf{y}_{J^C} - \mathbf{P}_{J^C}\mathbf{x})\|^2_2 + \mu \lVert\mathbf{x}-\mathbf{z}^{(k-1)}\rVert_{2}^2 \label{Eq:recons-vs-acgn-a}
    \\
    & \mathbf{z}^{(k)} = \arg \min_\mathbf{z}\mu \lVert \mathbf{x}^{(k)}-\mathbf{z}\rVert_{2}^2 +\cal{R}(\mathbf{z})\label{Eq:recons-vs-acgn-b}
\end{align}
\end{subequations}
where $\mu$ denotes the penalty parameter. Sub-problem \ref{Eq:recons-vs-acgn-a} can be solved efficiently using a conjugate gradient approach \cite{aggarwal2018modl}, and \ref{Eq:recons-vs-acgn-b} is solved by CNNs as before. 

In a separate illustrative experiment on the BSD68 dataset, structured noise is generated by drawing spatially correlated instances from a colored Gaussian distribution with a synthetic positive definite matrix as the covariance matrix. The noise is generated via the application of a $80 \times 80$ ideal band-pass filter on i.i.d. Gaussian noise in the discrete cosine transform space, where the total energy of the filter coefficients per pixel was constrained to be 100. For these experiments, we use the pre-trained regularizer CNNs from Section \ref{sec:bsd_experiments}, but modify the data fidelity term accordingly for N2I during inference time. 

Figure \ref{fig:colored_noise} shows that the proposed method N2I with a modified data fidelity term incorporates the statistical noise model successfully and removes the structured noise, whereas BM3D and all data-driven denoising methods (N2S/N2N/N2T) fail to remove the spatially correlated noise. Table \ref{tbl:structured_noise} summarizes the average PSNR values for the structured noise. This example highlights how Noise2Inpaint can readily be adapted to different statistical noise without re-training, owing to its ability to incorporate domain knowledge about noise and mask models. 

\section{Conclusions} \label{sec:conclusion}
We proposed the Noise2Inpaint approach, a self-supervised deep learning algorithm for image denoising from only noisy images. In particular, we first recast the denoising problem with holdout self-supervision as an iterative regularized inpainting problem consisting of data fidelity and regularizer terms, and then unroll the the iterative algorithm for fixed number of iterations. The training of this network was performed end-to-end by partitioning the noisy image pixels into two disjoint sets, similar to the purely data-driven Noise2Self, where one set was utilized in the data fidelity units of the unrolled network, while the other was used to define the loss. The experiments on different datasets showed that the proposed Noise2Inpaint outperforms its purely data-driven counterpart Noise2Self.

Furthermore, the objective function used for inpainting is able to incorporate different noise statistics both in training and testing. This was shown with colored Gaussian noise in Section \ref{sec:46}, but its application to non-Gaussian statistics is especially important for biomedical and biological applications, such as microscopy and MRI, where acquisition of clean target is often challenging and noisy data may be corrupted with non-Gaussian or colored noise.

{\small
\bibliographystyle{ieee_fullname}
\bibliography{egbib}
}

\end{document}